\documentclass[10pt,preprint]{aastex}
\usepackage{amsmath}
\usepackage{graphicx}
\usepackage{graphics}
\usepackage[dvips]{color}
\def\J1450{\mbox{SDSS~J1450+5300}}

\shorttitle{SDSS J1450+5300}
\shortauthors{Zhang et al.}

\begin{document}

\title{Ultra-dense Broad-line Region Scale Outflow in Highly Reddened Quasar SDSS J145057.28+530007.6}
\author{Shaohua Zhang\altaffilmark{1}, Hongyan Zhou\altaffilmark{1,2},Xiheng Shi\altaffilmark{1}, Xiang Pan\altaffilmark{1,2}, Tuo Ji\altaffilmark{1} and Peng Jiang\altaffilmark{1}}
\affil{$^1$SOA Key Laboratory for Polar Science, Polar Research Institute of China, 451 Jinqiao Road, Shanghai, 200136, China; zhangshaohua@pric.org.cn, zhouhongyan@pric.org.cn\\
$^2$CAS Key Laboratory for Research in Galaxies and Cosmology, Department of Astronomy,
    University of Sciences and Technology of China, Hefei, Anhui 230026, China}

\begin{abstract}
We report the discovery of highly reddening and hydrogen Balmer and metastable helium broad absorption lines in the quasar SDSS J145057.28+530007.6,
based on the optical and near-infrared spectra taken from the SDSS-III/BOSS and the  TripleSpec observations.
The nuclear continuum, Balmer decrement and absorption-line depth analyses suggest that (1) the accretion disk is completely obscured
and the covering factor of the broad-line region (BLR) is only $0.39\pm0.03$,
(2) the power-law continuum is reddened by the SMC extinction law of $E(B-V )=0.72\pm 0.01$ mag
and the dusty materials are mainly associated with \ion{Ca}{2} H and K rather than the Balmer and \ion{He}{1}* absorption-line system,
(3) the unsaturated  Balmer (H$\beta$, H$\gamma$, and H$\delta$) and \ion{He}{1}* $\lambda3889$ absorption lines have same two-Gaussian profiles with the shifts of $-931\pm33$ and $-499\pm39$ km s$^{-1}$ and the widths of $121\pm28$ and $196\pm37$ km s$^{-1}$, respectively.
Constrained mutually by the Balmer, \ion{He}{1}* absorption lines and undetected \ion{Fe}{2}* $\lambda5169$ in the photoionization simulations,
the physical properties of the outflow gases are derived as follows:
ionization parameter $10^{-1.4} \lesssim U \lesssim 10^{-0.8}$,
density $10^{8.2\pm0.4} \lesssim n_{\rm H} \lesssim 10^{9.0\pm0.4}$ cm$^{-3}$,
and column density $10^{22.0\pm0.2} \lesssim N_{\rm H} \lesssim 10^{22.2-22.3}$ cm$^{-2}$.
We propose that the ultra-dense outflow gases appear in the vicinity of the surface of the BLR
or are located at most 3.12 pc away from the engine.
That probably implies that the outflow originates from the BLR, and this kind of ultra-dense BLR scale outflow gases
can effectively test the physical properties of the outer gases of the BLR.

\end{abstract}

\keywords{galaxies: active -- quasars: absorption lines -- quasars: individual (SDSS J145057.28+530007.6)}
\maketitle

\section{Introduction}
In early of the twentieth century, astronomers confirmed that our Milky Way galaxy is in reality just one of hundreds of billions of galaxies in the universe. Now we know that each massive galaxy consists of billions of stars, myriad clouds of gas and dust, and a super-massive black hole (SMBH) at its center.
There are still some challenges remaining to be solved, such as ``How do black holes grow, radiate, and influence their surroundings?'' and ``How do baryons cycle in and out of galaxies, and what do they do while they are there?''\footnote{Two key questions are identified by ``New Worlds, New Horizons in Astronomy and Astrophysics''. The book can be downloaded at http://sites.nationalacademies.org/bpa/bpa\_049810 .}, which are closely related to the topics of the SMBHs and host galaxies co-evolution (e.g., Granato et al. 2004; Scannapieco \& Oh 2004; Hopkins et al. 2008), and the properties of the gas clouds in the structures of active galactic nuclei (AGN) (e.g., Ferland \& Osterbrock 1986; Peterson 1993; Netzer \& Peterson 1997; Wang et al. 2013).

The blueshifted emission lines and blueshifted/redshifted intrinsic absorption lines are once believed to be an useful approach to diagnosing properties of outflowing (feedback to the host, e.g., Gaskell 1982; Weymann et al. 1991) or inflowing (SMBH accretion, Shi et al. 2016b) gases at different scales.
However, the blueshifted emission lines are generally mixed with the normal emission lines (e.g., Komossa et al. 2008; Zhang et al. 2011; Wang et al. 2011; Marziani et al. 2013; Liu et al. 2016), with three exceptions in which the blueshifted emission lines dominate the emission profiles (IRAS 13224-3809 and 1H 0707-495, Leighly \& Moore 2004; SDSS J000610.67+121501.2, Zhang et al. 2017a).
Meanwhile, the ubiquitous broad absorption lines (e.g., \ion{C}{4} and \ion{Mg}{2} BALs) are commonly blending and saturated (e.g., Hall et al. 2002; Trump et al. 2006; Gibson et al. 2009; Zhang et al. 2010; Allen et al. 2011),
and their variations can only assess the kinetic properties (e.g., Hall et al. 2011; Zhang et al.  2015a; Rafiee et al. 2016) or whether variations of the ionization of gas (e.g., Hamann et al. 2008, Filiz Ak et al. 2013; Trevese et al. 2013; Wang et al. 2015; He et al. 2017). Thus they are difficult to be used to accurately study the physical condition and geometry of the gas winds.

The detection of absorption lines from hydrogen Balmer and metastable helium in NGC 4151 (Anderson 1974) rekindled the hope for the quantitative study of AGN absorption lines. (1) Both ions have multiple upward transitions in a wide wavelength span from the ultraviolet (UV) to the near-infrared (NIR), which are easy to be observed. (2) There is no blending problem since the transitions are widely separated. (3) Multiple lines from the same lower level are very helpful in jointly determining the column density and covering factor of the lines (e.g., Arav et al. 2005). Meanwhile, $\rm H(n=2)$ is sensitive to the gas density, while \ion{He}{1}* is sensitive to the ionization state of the absorption gases (e.g., Arav et al. 2001; Ji et al. 2015). Thus the absorptions of hydrogen Balmer and \ion{He}{1}* can provide us abundant information about the absorption gases, such as velocity distribution, (column) density, ionization state, and furthermore distance from the ionization source and even kinetic energy and mass-flow rate.
For that reason, samples of \ion{He}{1}* BALs (Liu et al. 2015), \ion{He}{1}* NALs (T. Ji et al. in preparation) and hydrogen Balmer absorption lines (X.-H. Shi et al. in preparation) have been explored in the Sloan Digital Sky Survey (SDSS, York et al. 2000) -I/II and -III quasar's spectral databases (Schneider et al. 2010; P{\^a}ris et al. 2017) as well as the detailed studies for a few individuals (e.g., Aoki et al. 2006, 2010; Hall et al. 2007; Arav et al. 2008; Leighly et al. 2011, 2014, 2015; Zhang et al. 2015b, 2017b; Shi et al. 2016a, 2016b, 2017; Sun et al. 2017).

In this work, we report on a quasar (SDSS J145057.28+530007.6, hereafter \J1450) with an emission redshift of $z_{\rm emi}=0.9166\pm0.0001$. \J1450\ shows the broad absorption troughs of hydrogen Balmer (from H$\alpha$ to H$\delta$) and metastable helium (\ion{He}{1}* $\lambda\lambda$3889, 10830) with the width of $\sim1200$ km $s^{-1}$, suggesting the ultra-dense broad-line region (BLR) scale outflow materials in the nuclear region. \J1450\ also is the fifth Balmer BAL quasar after SDSS J125942.80+121312.6 (Hall 2007; Shi et al. 2016a), LBQS 1206+1052 (Ji et al. 2012), SDSS J222024.59+010931.2 (Ji et al. 2013), and SDSS J152350.42+391405.2 (Zhang et al. 2015b).
The data we used will be described in Section 2. We will analyze the reddened continuum and absorption lines in Section 3, and discuss the properties and possible origins of  outflows in Section 4. A summary of our results will be given in Section 5. Throughout this paper, we adopt the cold dark matter `concordance' cosmology with H$_{\rm 0}$ = 70 km s$^{-1}$Mpc$^{-1}$, $\Omega_{\rm m} = 0.3$, and $\Omega_{\Lambda} = 0.7$.

\section{Observations}
\J1450\ is an infrared-luminous dust-reddened quasar. The broad-band spectral energy distributions (SEDs) from the ultraviolet out to the infrared  are presented by the photometric images taken with the SDSS at $u$, $g$, $r$, $i$, and $z$ bands,
the Two Micron All Sky Survey (2MASS; Skrutskie et al. 2006) at $J$, $H$, and $K_s$ bands,
and the Wide-field Infrared Survey Explorer (WISE; Wright et al. 2010) at $W1$, $W2$, $W3$, and $W4$ bands. The optical-NIR color-color diagram represents that \J1450\ just meets at the selection criteria of the FIRST-2MASS red quasar in Glikman et al. (2007). The multi-band magnitudes are summarized in Table \ref{tab1}.

The optical spectrum of \J1450\ was taken with the SDSS 2.5 m telescope on April 8, 2013, in the SDSS-III Baryon Oscillation Spectroscopic Survey (BOSS; Dawson et al. 2013)
and published in the SDSS Twelfth Data Release (DR12; Alam et al. 2015). The BOSS spectrum has a wider wavelength range covering 361 nm - 1014 nm with a resolution of
1300 at the blue side and 2600 at the red side. There are strong broad/narrow emission lines of H$\beta$, H$\gamma$, [\ion{O}{3}] and [\ion{O}{2}] doublets, and \ion{Mg}{2}, etc, and abundant remarkable absorption lines of hydrogen Balmer series,  \ion{He}{1}* $\lambda3889$, and \ion{Ca}{2} H and K, in the ``not-high'' S/N ratio spectrum.

Two years later, the NIR spectrum of \J1450\ was performed with the TripleSpec spectrograph of the Hale 200-inch telescope (P200) at Palomar Observatory on May 26, 2015. Four exposures of 300 seconds each are taken in an A-B-B-A dithering model.
TripleSpec (Wilson et al. 2004) provides simultaneous wavelength coverage from 0.9 to 2.46 $\mu$m at a resolution of 1.4 - 2.9 \AA\ with two gaps at approximately 1.35 and 1.85 $\mu$m owing to the telluric absorption bands.
The raw data were processed using IDL-based Spextool software (Vacca et al. 2003; Cushing et al. 2004).
The H$\alpha$ emission/absorption lines and \ion{He}{1}* $\lambda10830$ absorption line are detected with the TripleSpec at $J$ band and $K_s$ band, respectively.

\section{Data Analysis}

\subsection{Reddened Nuclear Continuum}

\begin{figure*}[tbp]
\epsscale{1.0} \plotone{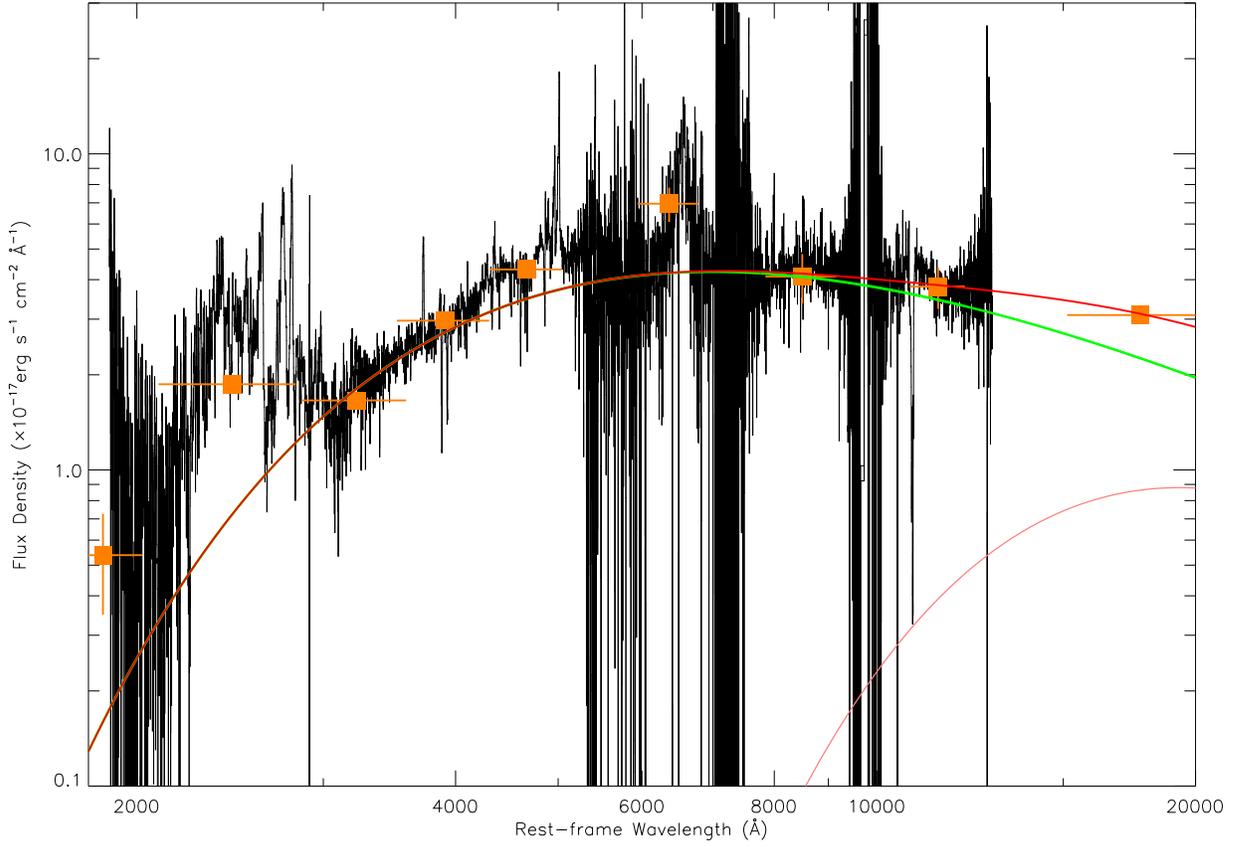}
\caption{Broad-band SED of SDSS~J1450+5300 from UV to NIR by brown squares, the spectra of SDSS and TripleSpec by black curves.
The reddened power-law ($\propto \lambda^{-1.7}$) continuum with $E(B-V)=0.72$ mag, the hot ($T=1500$ K) dust emission and their sum are shown by green, pink and red curves. }\label{f1}
\end{figure*}

In Figure \ref{f1}, we plotted the broad-band SED of \J1450\ by brown squares in log-space.
The multi-band magnitudes are firstly corrected for the Galactic reddening of $E(B-V)=0.017$ mag (Schlegel et al. 1998), and then transformed to the quasar's rest frame. Meanwhile, the optical and NIR spectroscopic data are scaled to match the SDSS and UKIDSS photometry, and overplotted in the figure by black curves.
The arch structure from $\sim 3000$ \AA\ to 2$\mu$m implies that the nuclear continuum is extremely reddened.
As discussed in Zhang et al. (2017b), variability of the continuum strength and shape could be an issue in \J1450\ because of the large time interval (8.8 yr in the rest frame) between the observations. However, the perfect matching of the optical/NIR spectra simply multiplied by a factor and the broad-band SED rule out the possibility of variability affecting on the SED.

Because of the pollution of H$\alpha$ and H$\beta$ broad lines, and optical \ion{Fe}{2} multiplets,
we just chose the magnitudes at $r$, $i$, $H$, $K_s$, and $W1$ bands without strong emission lines to constrain the continuum.
The phenomenological model contains two components: (1) a reddened power-law continuum from the accretion disk using the SMC-type extinction law (Pei 1992), and (2) a hot dust emission from the dusty torus.
Based on the above analysis, we can decompose the broad-band SED (3000 - 20000 \AA\ in rest frame wavelength) with the following model:
\begin{align}
F_{\lambda} = C_{nuclear}~A_{nuclear}\left(E(B-V),\lambda\right)~\lambda^{\alpha} + C_{bb}~B_{\lambda}\left(T_{dust}\right),
\label{eq1}
\end{align}
where $C_{nuclear}$ and $C_{bb}$ are the factors for the respective components, $A\left(E(B-V),\lambda\right)$ is the dust extinction to the power-law continuum,
and $B_{\lambda}\left(T_{dust}\right)$ is the Planck function.
As the continuum slope and extinction are somehow degenerate, here $\alpha$ is fixed to $-1.7$ (the mean value of the quasar UV/optical continuum slope),
which is the common recipe for reddened AGN continua in the literature (see, e.g., Dong et al. 2005, 2012, and Zhou et al. 2006).
In addition, $T_{dust}$ is set to the dust sublimation temperature $T_S =1500\rm~ K$, which it is  as the typical temperature of the hot dust in torus (e.g., Sanders et al. 1989).
Thus, there are three free parameters, i.e., $E(B-V)$, $C_{nuclear}$ and $C_{bb}$, in the fitting.
We perform least-squares minimization using the Interactive Data Language (IDL) procedure MPFIT (Markwardt et al. 2009), the value of $E(B-V)$ is $0.72\pm0.01$ mag.

In Figure \ref{f1}, the best-fit model (red curve) is  in good agreement with the continuum windows from the optical out to NIR wavelengths.
The reddened power-law continuum and hot dust emission are also shown by green and pink curves.
It should be noted that the SDSS spectrum has the``continuous'' excess fluxes with the wavelength of $\lambda < 3000$ \AA.
%Except for the normal strong UV \ion{Fe}{2} multiplets (can be modeled using the \ion{Fe}{2} template from Vestergaard \& Wilkes 2001), careful study will find the extra UV \ion{Fe}{2} emission components peak approximately at 2620 and 2745 \AA.
The unusual \ion{Fe}{2} emission gives an illusion of absorption troughs in the wavelength ranges of $\sim 2630-2730$ \AA\ and $\sim 2755-2790$ \AA.
The analysis and discussion about the \ion{Fe}{2} multiplets origin will be present in P. Jiang et al. (in preparation),
this work will focus on the absorption-line system of hydrogen Balmer and \ion{He}{1}*.

\subsection{Emission Lines}
It becomes apparent that the ``naked'' eye can find the absorption of hydrogen Balmer  and  \ion{He}{1}*.
The measurement of the absorption strengths is generally by through comparing the observed spectrum and the unabsorbed template.
For the broad absorption troughs, there are two methods generally used to obtain the the unabsorbed template, i.e., the ``pair-matching'' method (e.g., Zhang et al. 2014; Liu et al. 2015), and the spectral-decomposition method (for \ion{C}{4}, Gibson et al. 2008; for \ion{Mg}{2}, Zhang et al. 2010; for H$\alpha$ and H$\beta$, Zhang et al. 2015b).
Pair-matching method is much more effective for the weak and shallow absorption troughs, and the seriously mutilated emission-line profiles which are very hard to process spectral decomposition. However, the weakness of the pair-matching method is all spectral components (continuum, \ion{Fe}{2} multiplets, and broad emission lines) must have one unity covering factor, the spectral-decomposition method is more flexible for the absorption-line measurements.

%To further study the absorption lines, we firstly need to obtain the unabsorbed template of \J1450.
It is essential that the unabsorbed broad-line profiles are determined through the spectral-decomposition method.
The train of thought is briefly as follow:
(1) The nuclear continuum and hot dust emission adopt the best-fitting result of the broad-band SED.
(2) The strength, shift and width of the optical \ion{Fe}{2} multiplets will be ascertained in the wavelength range of \mbox{4000 - 5400} \AA\ (H$\beta$, H$\gamma$, H$\delta$, and [\ion{O}{3}] doublet are masked), which are also used for the \ion{Fe}{2} emission under H$\alpha$.
(3) Multi-Gaussian profile is implied to model the isolated and uncomplicated H$\alpha$ line.
(4) The emission lines of H$\beta$, H$\gamma$, and H$\delta$ are depicted by the scaled H$\alpha$ profile.
(5) The emission lines of \ion{He}{1} $\lambda10830$ and [\ion{Ne}{3}] 3868 are directly described by one single Gaussian curve.

\begin{figure*}[tbp]
\epsscale{1.0} \plotone{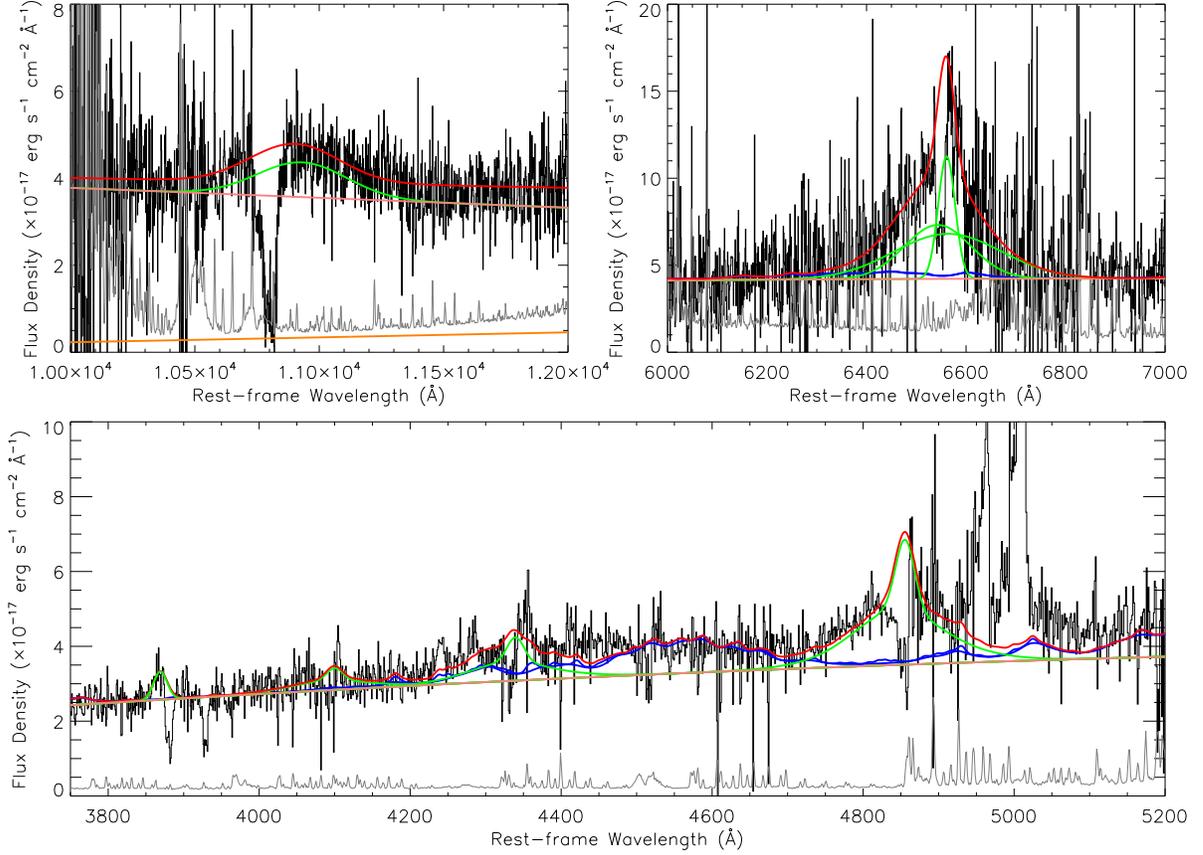}
\caption{Top-left: the TripleSpec NIR spectrum (black) and best-fit model (red) of \ion{He}{1} $\lambda10830$ regime. The reddened power-law continuum and hot dust emission from SED fitting and a gaussian broad emission-line profile are shown in pink, orange and green.
Top-right: the TripleSpec NIR spectrum (black) and best-fit model (red) of H$\alpha$ regime. The reddened power-law, broadened optical \ion{Fe}{2} template, and three gaussians of H$\alpha$ broad-line are shown by pink, blue, and green curves.
Bottom: the SDSS spectrum (black) and best-fit model (red) of H$\beta$-\ion{He}{1} $\lambda3889$ regime. We apply the three-Gaussian profile of H$\alpha$ to model the emission profiles of H$\beta$, H$\gamma$, and H$\delta$, and a singel Gaussion to model [\ion{Ne}{3}] 3868.
The reddened power-law, broadened optical \ion{Fe}{2} template, and emission profiles are shown by pink, blue, and green curves.
}\label{f2}
\end{figure*}

In step (2), we adopt the I Zw 1 \ion{Fe}{2} template provided by V{\'e}ron-Cetty et al. (2004) and convolve it with a Gaussian kernel in velocity space to match the width of \ion{Fe}{2} multiplets in the observed spectrum. Ther are six free parameters, i.e., the strength, shift, and width for broad and narrow \ion{Fe}{2} components, respectively.
The broadened \ion{Fe}{2} template and the nuclear continuum join to form the so-called the ``pseudo-continuum''.
We subtracted the pseudo-continuum from the observed spectra, and obtained the emission-line spectra of H$\alpha$ and H$\beta$.
In the next steps, we just mask the potential absorption regions in the emission-line fitting.
In the panels of Figure \ref{f2}, we plot the observed optical/NIR fluxes and errors by black and gray curves. The reddened power-law continuum from the broad-band SED fitting, broadened optical Fe II template are shown in pink and blue. The green curves represent the Gaussian components of H$\alpha$, the whole profiles of H$\beta$, H$\gamma$, and H$\delta$, and also the single Gaussian profiles of \ion{He}{1} $\lambda10830$ and [\ion{Ne}{3}] $\lambda3868$. The sum of pseudo-continuum and emission lines are overplotted by red curve.

\subsection{Absorption lines}

After obtaining the emission-line spectrum, we will try to study the situation of gas absorption step by step.
Let's first look at the top-left panel of Figure \ref{f2}, the bottom of the \ion{He}{1}* $\lambda10830$ absorption trough has reached the hot dust emission (orange curve), that implies the absorption gases completely obscure the accretion disk -- the continuum source.
Do the absorption gases completely or partially cover the BLR? The absorption troughs of hydrogen Balmer series can clarify the question.

\begin{figure*} [htbp]
\epsscale{1.0} \plotone{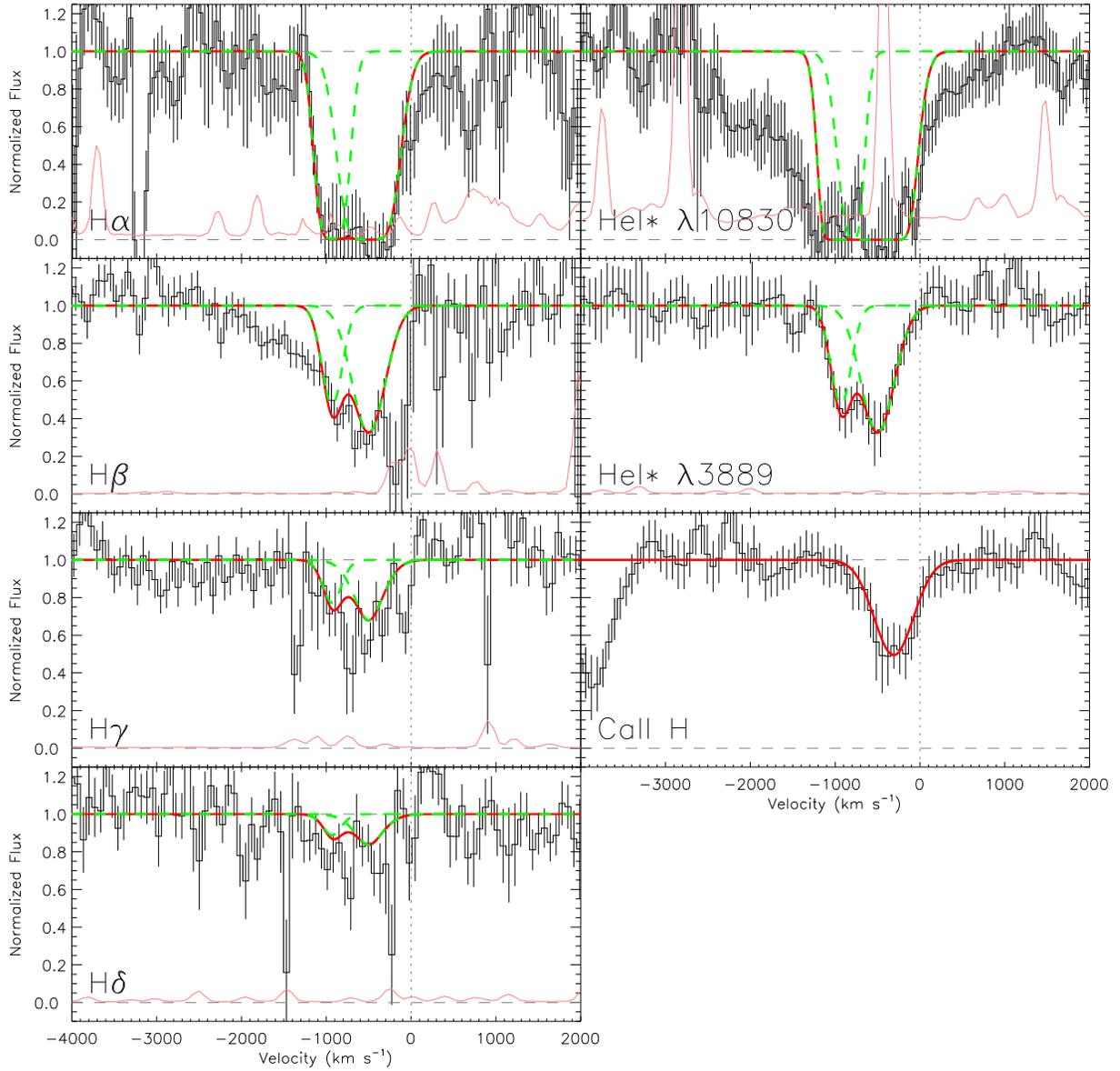}
\caption{Left: the normalized absorption profiles for H$\alpha$, H$\beta$, H$\gamma$, and H$\delta$ from top to bottom.
Right: the normalized absorption profiles for \ion{He}{1}* $\lambda10830$, \ion{He}{1}* $\lambda3889$, and \ion{Ca}{2} $\lambda3933$.
The theoretical double-Gaussian absorption troughs for hydrogen Balmer lines and \ion{He}{1}* $\lambda3889,10830$  are overploted in red, and the sky line spectrum (not scaled) is shown as a pink line in the panels.
}\label{f3}
\end{figure*}

Based on the fact of the accretion disk completely obscured, for a given covering factor $Cf$ of the absorption gases to the emission lines (broad emission lines and optical \ion{Fe}{2} multiplets), the absorption-line spectrum (the normalized intensities) is
\begin{align}
I_{\lambda} = \dfrac{F_{\lambda} - (1-Cf)~F_{\lambda, ELine}}{F_{\lambda, Conti} + Cf~F_{\lambda, ELine}},
\label{eq2}
\end{align}
where $F_{\lambda}$, $F_{\lambda, ELine}$, and $F_{\lambda, Conti}$ are the observed spectrum, the emission-line spectrum, and the nuclear continuum, respectively.
Here, the absorption gases are assumed to have a unity covering factor in different velocities.
The residual fluxes in the H$\alpha$ trough are almost two times than those in the H$\beta$ trough, the BLR is partially obscured.
Meanwhile, the troughs of \ion{He}{1}* $\lambda10830$ and H$\alpha$ look like flat-bottom, we guess that they probably are saturated.
In this case, the covering factor is $Cf=0.39\pm0.03$.
We also traverse the parameter space ($0.39 \le Cf < 1.0$) to search the available covering factor,
which can ensure the absorption depths of the H$\alpha$, H$\beta$, and H$\gamma$ troughs matching with their known oscillator strengths.
The attempt confirmed that the absorption of H$\alpha$ is saturated rather than other Balmer lines.
We adopted the above covering factor, calculated the normalized fluxes for hydrogen Balmer series and \ion{He}{1}* $\lambda\lambda$3889,10830,
and presented their absorption troughs in velocity space in Figure \ref{f3}.
Overall, the absorption troughs in \J1450\ are BALs (or mini-BALs), for example, the width of the H$\alpha$ trough, with the observed spectrum falling at least 10\% below the unabsorbed model, is $ 1214\rm ~km~s^{-1}$.

Panels of Figure \ref{f3} show that the \ion{He}{1}* and hydrogen Balmer absorption lines have similar profiles, thus, we assume
the \ion{He}{1}* and hydrogen Balmer absorbers share the same kinematic structure and can fit two Gaussians to the normalized fluxes of these lines. We firstly started with \ion{He}{1}* $\lambda3889$, since the \ion{He}{1}* $\lambda3889$ trough is clear and neat, and has the highest S/N ratio. The velocity shifts with respect to the quasar's rest frame are $-931\pm33$ km s$^{-1}$ and $-499\pm39$ km s$^{-1}$ for the two components respectively. The negative value means the absorption-lines are blueshifted. The widths are $121\pm28$ km s$^{-1}$ for the high-velocity component and $196\pm37$ km s$^{-1}$ for the low-velocity component.
 One can find that the sum of two Gaussians can reconstruct the transmission of \ion{He}{1}* $\lambda3889$ very well. 
Secondly, we used two-Gaussian profiles with the same shifts and widths as those of \ion{He}{1}* $\lambda3889$ to model the H$\beta$ trough, the depths of the two components are free and are contained by the bottom of the trough. In the panels of \ion{He}{1}* $\lambda3889$ and H$\beta$, the Gaussian components and their sum are shown by green and red curves.

The true optical depth ($\tau_v$) as functions of radial velocity for the relevant ion is $\tau_v = - ln~ I_v$,
and then, the column densities on the hydrogen $n=2$ shell and metastable $\rm He^0~2^3S$ as a function of velocity are calculated using the general expression (e.g., Arav et al. 2001)
\begin{align}
N(\Delta v) = \dfrac{3.7679\times10^{14}}{\lambda_0 f_{ik}} \tau(\Delta v) \rm~ [cm^{-2}~(km ~s^{-1})^{-1}],
\label{eq3}
\end{align}
where $\lambda_0=4862.68$ \AA\ and $3889.80$ \AA\ are the wavelengths of the H$\beta$ and \ion{He}{1}* $\lambda3889$ lines and $f_{ik}=0.1190$ and $0.0644$ are the corresponding oscillator strengths\footnote{Oscillator strengths are from NIST Atomic Spectra Database (http://physics.nist.gov/PhysRefData/ASD/)}, respectively. We obtain the total  column densities $N_{\rm H(n=2)}=(4.89\pm1.07)\times10^{14}\rm~ cm^{-2}$ and $N_{\rm HeI*}=(1.01\pm0.19)\times10^{15}\rm~cm^{-2}$ by integrating Equation \ref{eq4}. The theoretical absorption profiles of H$\alpha$, H$\gamma$, H$\delta$, and \ion{He}{1}* $\lambda10830$, obtained from the column densities as a function of velocity, are shown in the panels of Figure \ref{f3}.

\section{Discussion}

From the emission-line fittings, we can derive the central black hole mass using the commonly used virial mass estimator. We use the $L_{\rm 5100}$ and $FWHM_{\rm H\beta}$ based mass formalism given by Greene \& Ho (2005), $M_{\rm BH}=(4.4\pm0.2) \times 10^6~ \left(\dfrac{L_{\rm 5100}}{10^{44}~{\rm erg~s^{-1}}}\right)^{0.64\pm0.02} \left(\dfrac{FWHM_{\rm H\beta}}{10^3~{\rm km~s^{-1}}}\right)^{2} M_{\sun}$.   $L_{\rm 5100}$ and $FWHM_{\rm H\beta}$ are the luminosity and full width at half maximum of H$\beta$ line.
The monochromatic continuum luminosity $L_{\rm 5100}\left(=\lambda L_{\lambda}(\rm 5100\AA)\right)=1.51\times10^{45}$ \mbox{erg~ s$^{-1}$} at 5100 \AA\ is directly calculated from the reddened power-law continuum, and the extinction corrected luminosity $L_{\rm 5100}=1.19\times10^{46}$ \mbox{erg~ s$^{-1}$}.
Together with $FWHM_{\rm H\beta}\sim FWHM\rm_{H\alpha}=3011~km~ s^{-1}$ (H$\beta$ has the same profile as H$\alpha$ in velocity space),
the central black hole mass is estimated to be $M\rm_{BH}=8.50\times10^8~\rm M_{\sun}$.
Ho \& Kim (2015) suggested that intrinsic scatter of the single-epoch black-hole mass estimate method is $\sim 0.41$ dex, then the central black hole mass has an uncertainty of a factor of $\sim 2.6$.

The bolometric luminosity is estimated from the monochromatic luminosity using the conversion given by Runnoe et al. (2012),
 then $L_{\rm bol} = 0.75 \times 10^{~(4.89\pm1.66)+(0.91\pm0.04)~ {\rm log_{10}} L_{\rm 5100}}=6.08\times10^{46}$
  \mbox{erg~s$^{-1}$}.
In Zhang et al. (2015b), the bootstrap approach showed the uncertainty of $L_{\rm bol}$ is in the order of ten percent (also see Dong et al. 2008). The derived Eddington ratio is thus $l_{\rm E}=L_{\rm bol}/L_{\rm Edd}=0.57$.
Based on the bolometric luminosity, the amount of mass being accreted is estimated as
$\dot{M}_{acc}=L\rm_{bol}/\eta c^2 =10~M_{\sun}~yr^{-1}$,
where we assumed an accretion efficiency $\eta$ of 0.1, and $c$ is the speed of light.
The radius of broad-line region, $R_{\rm BLR}$,
can be estimated using the formula based on the luminosity at 5100\AA,
 $R_{\rm BLR} = \alpha \left(L_{5100}/10^{44}~\rm erg~s^{-1}\right)^{\beta}$  lt-days.
 The parameters, $\alpha$ and $\beta$ are $30.2\pm1.4$ and $0.64\pm0.02$ given in Greene \& Ho (2005) and $20.0^{+2.8}_{-2.4}$ and $0.67\pm0.07$ given in Kaspi et al. (2005), respectively.
Thus, the luminosity yields $R_{\rm BLR}=0.54-0.60$ pc.
Meanwhile, the radius of the inner side of the dusty torus (the dust sublimation radius), $R_{\rm Torus}$, can also be estimated based on the thermal equilibrium as $ R_{\rm Torus}=\sqrt{L_{\rm bol}/4\pi\sigma T^4}$,
where $\sigma$ is the Stefan-Boltzmann constant, $T (\sim 1500~\rm K)$ is the temperature of inner side of the tours (Barvainis 1987).
Then, we get $R_{\rm Torus} = 8.56$ pc.
Similarly, the extend scales of the torus are on the scale of $\rm 10$ pc (Burtscher et al. 2013; Kishimoto et al. 2011).

%\subsection{Physical properties of the absorption gases}

To investigate the physical properties for the absorption gases, we use the  photoionization synthesis code Cloudy ({\bf V16.0}, last described by Ferland et al. 1998) simulations
and confront these models with the measured column densities of ions to determine the density ($n\rm_H$), the column density ($N\rm_H$) and the ionization parameter ($U$).
We consider a gas slab illuminated by a continuum source in the extensive parameter space.
The absorption gases are assumed to have unity density and a homogeneous chemical composition of solar values and be free of dust.
The incident SED applied is a typical AGN multi-component continuum described as a combination of a blackbody ``Big Bump''
and power laws\footnote{see details in Hazy, a brief introduction to $Cloudy$; http://www.nublado.org}.
We calculated a series of photoionization models with different ionization parameters, electron densities and
hydrogen column densities. The ranges of parameters are $-2\leqslant{\rm log_{10}}~U\leqslant0$,
$4\leqslant{\rm log_{10}}~n_{\rm H}~({\rm cm^{-3}})\leqslant13$ and
$20\leqslant{\rm log_{10}}~N_{\rm H}~({\rm cm^{-2}})\leqslant24$ with a step of 0.1 dex.

\begin{figure} [h]
\centering
\rotatebox{270}{\includegraphics[width=0.78\textwidth]{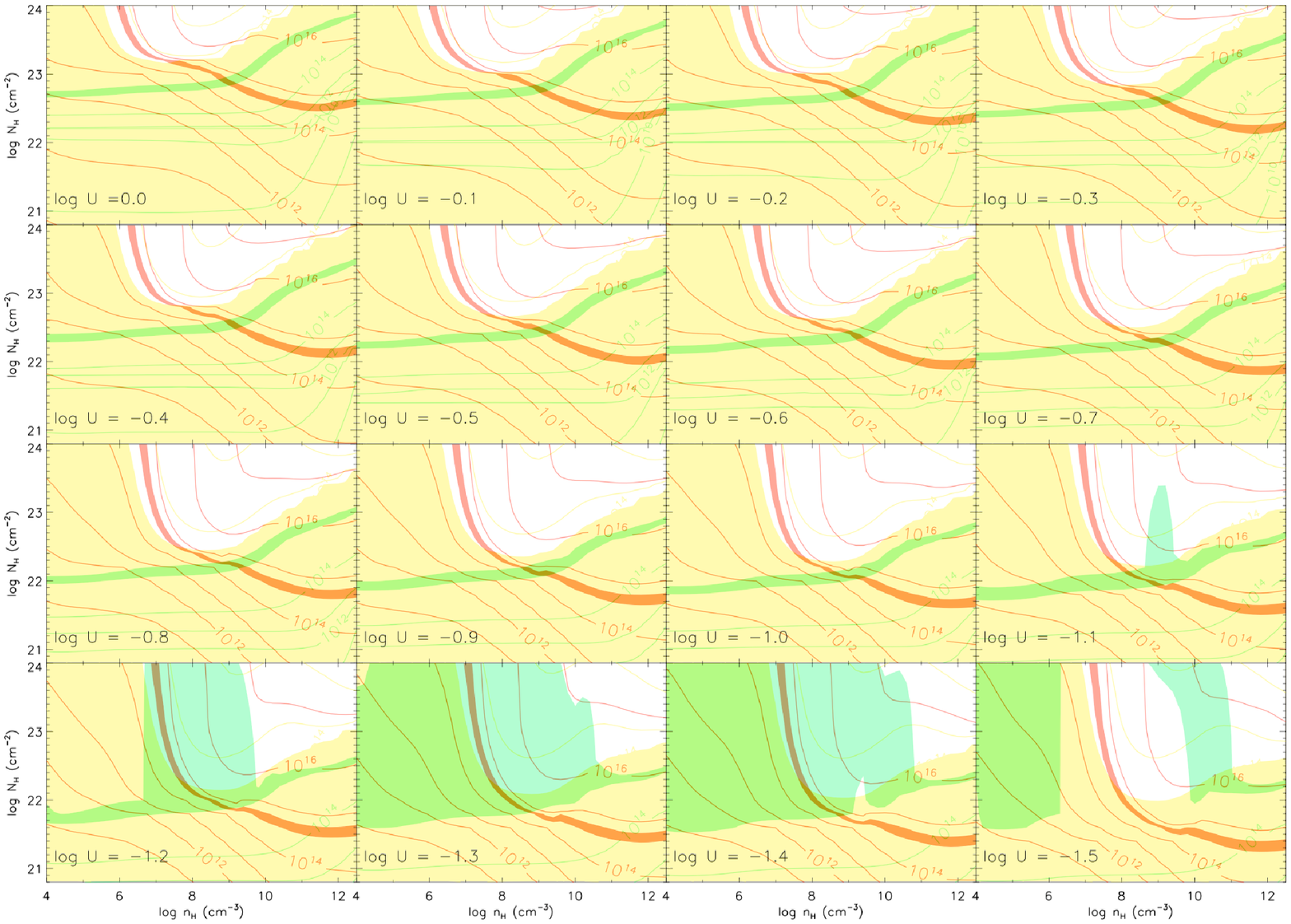}}
\caption{Predicted ionic column densities from the photoionization simulations by Cloudy with $U$ from $10^{0.0}$ to $10^{-1.5}$ as functions of $n_{\rm H}$ and $N_{\rm H}$.
The attached numbers are logarithms of ionic column densities. The red, green, and yellow contours show the {\bf ions} of $\rm H^0_{n=2}$, $\rm He^0~2^3S$, and $\rm Fe^+~{_a^6S_{5/2}}$,
and their measured column densities are overploted by the corresponding colored areas.
%The model of $U \geqslant 10^{-1.4}$, $n_{\rm H} \approx  10^{8-9}\rm ~cm^{-3}$, and $N_{\rm H} = 10^{21.8-23}\rm ~cm^{-2}$ can reproduce the measured strengths of Balmer lines, \ion{He}{1}* and optical \ion{Fe}{2} simultaneously.
}\label{f4}
\end{figure}

We extract the resultant column densities on the hydrogen $n=2$ shell and metastable $\rm He^0~2^3S$
from the Cloudy simulations and show them by the red and green lines in Figure \ref{f4}.
 The red and green areas show the observed $1\,\sigma$ uncertainty ranges of $N\rm_{H(n=2)}$ and $N\rm_{HeI*}$,
so the overlapping region is the possible parameter space ($U > 10^{-1.4}$, $n\rm _{H}\sim 10^{6-9}~cm^{-3}$, $N\rm_{H} > 10^{21.8}~cm^{-2}$) for the absorption gases of \J1450.
 Shi et al. (2016a) used the isolated optical \ion{Fe}{2} lines further to narrow the possible parameter space.
The undetected \ion{Fe}{2}* $\lambda5169$ gave that the upper limit of the column density is $\rm 3.37\times10^{14}~cm^{-2}$  for $\rm Fe^{+}~a^6 S_{5/2}$ level (yellow area in Figure \ref{f4}), otherwise, the \ion{Fe}{2}* $\lambda5169$ absorption trough will be detected at $2\,\sigma$ level. We find that the addition of \ion{Fe}{2}* $\lambda5169$ largely compress the parameter space of the (column) density in the range of  $U\sim10^{-1.4~-~ -1.2}$, however, it is no help to constrain the ionization parameter $U$.
The only valid limit for the ionization parameter still is $U > 10^{-1.4}$. 
If we adopted $0.3 \leqslant U \leqslant 10$ as the ionization parameter range of the BLR (Netzer 1993),
the upper limit of the absorption gases should be less than 0.3.
In fact, the ionization parameter cannot reach such a high value.
Based on the distance estimation of the absorption gases in the next paragraph, the gas winds have be located on the surface of the BLR ($R\sim 0.54-0.6 \rm~pc$) with the ionization parameter of $U\sim10^{-0.8}$.
 Thus, the value of $U\sim10^{-0.8}$ is used as the upper limit of the ionization parameter of the absorption gases.
%Finally, model calculations suggest that the gas clouds in the regions with $10^{-1.4}\lesssim U \lesssim 10^{-0.8}$.
For the lowest ionization parameter of $U=10^{-1.4}$, the (column) densities are $n_{\rm H}=10^{8.2\pm0.4}\rm ~cm^{-3}$ and
$N_{\rm H} = 10^{22.0\pm0.2}\rm ~cm^{-2}$, and they will be $n_{\rm H} =  10^{9.0\pm0.4}\rm ~cm^{-3}$ and $ N_{\rm H} = 10^{22.2-22.3}\rm ~cm^{-2}$ when the ionization parameter is equal to the highest value of $U=10^{-0.8}$.

Based on $U$ and $n_{\rm H}$ determined by Cloudy, we estimate the distance ($R$) of the absorption gases away from the central source.
$U$ depends on $R$ and the rate of hydrogen-ionizing photons emitted by the central source $Q$, as follows,
$U =  Q/4\pi R^2 n_{\rm H}c$, in which, $c$ is the speed of light.
To determine the $Q$, we scale the AGN multi-component continuum to the extinction corrected flux of \J1450\ at 5100\AA\
and then integrate over the energy range $h\mu\geqslant 13.6$ eV. This yields $Q = 2.16 \times 10^{56}~\rm photons~s^{-1}$.
Using this $Q$ value together with the derived lower limit of the ionization parameter $U=10^{-1.4}$ and the density of $n\rm_{H}=10^{8.2\pm0.4}\rm~cm^{-3}$, the upper limit of $R$ can be derived to as $\sim 3.1$ pc.
With the increase in the ionization parameter, the absorption gases will gradually close (reach) to the surface of the BLR.
However, if the spectral S/N ratios of the \ion{Fe}{2}* $\lambda5169$ regime are increased by ten times {\bf in the possible future observation with more exposure time}, the upper limit of $N_{\rm FeII*}$ will be lowered, and the ionization parameter will be limited to $U \geqslant 10^{-1.2}$.
Thus we are more inclined to believe that the absorption gases in \J1450\ attach to the BLR or appear in the vicinity
of the surface of the BLR.
Among cases with Balmer and/or \ion{H}{1}* absorbers, the absorber in \J1450\ is the second-nearest from the the central engine except SDSS J152350.42+391405.2 (Zhang et al. 2015b), whose absorption gaes are located at a distance of $\sim 0.2$ pc (slightly farther than that of the BLR).
Other cases are spread about on the scale from parsec-scale (Shi et al. 2016a, 2016b; low-velocity component, Sun et al. 2017) to dozens of parsec (Zhang et al. 2017) and even hundreds of parsec (Ji et al. 2015; Leighly et al. 2014) or kiloparsec (high-velocity component, Sun et al. 2017). Meanwhile, the high density of $n\rm_{H}\sim 10^9 ~cm^{-3}$
in \J1450\ is also present in SDSS J152350.42+391405.2 (Zhang et al. 2015), SDSS J125942.80+121312.6 (Shi et al. 2016a), SDSS J112526.12+002901.3 (Shi et al. 2016b), and LBQS 1206+1052 (low-velocity component, Sun et al. 2017).
Interestingly, Kaastra et al. (2014) reported a fast long-lived outflow in NGC 5548. The clumpy stream of ionized gas blocks 90\% of the soft X-ray emission and causes deep BAL broad troughs and it is at a distance of only a few light days from the nucleus (likely originate from the accretion disk). We also notice the similar properties (highly reddened and the small distance between the absorber and the nucleus) in \J1450, the absorber in \J1450\ may be an another case. Possible high-resolution X-ray and UV observations in the future would tell the similarities and differences between them.

Assuming that the absorption materials can be described as a thin partially filled shell,
the average mass-flow rate ($\dot{M}$) and kinetic luminosity ($\dot{E}_k$) can be derived as
$\dot{M}=4\pi R\Omega\mu m_p N_H v$ and $\dot{E}_k=2\pi R\Omega\mu m_p N_H v^3$ (Borguet et al. 2012),
where $R$ is the distance of the outflows from the central source, $\Omega$ is the global covering fraction of the outflows,
$\mu = 1.4$ is the mean atomic mass per proton, $m_p$ is the mass of proton,
$N\rm_H$ is the total hydrogen column density directly derived from the photoionization modeling of the outflow gases,
and $v$ is the flux weight-averaged velocity of the absorption gases.
Here, we adopted $R\approx 0.6-3.1$ pc, $\Omega(\equiv  Cf)=0.39$, $N\rm_H=10^{21.8-22.3}$ cm$^{-2}$, and $v\approx 600$ \mbox{km s$^{-1}$},
then the kinetic luminosity and mass loss rate are calculated as
$\dot{E_k} = 0.4-2.3\times10^{41}~\rm erg~s^{-1}$ and $\dot{M} = 0.7-11.6~\rm M_{\sun}~yr^{-1}$.
The kinetic luminosity of \J1450\  is just able to meet the reported lower limit on the kinetic luminosity ($\dot{E_k}\sim10^{-5}L_{Edd}$)
(e.g., Scannapieco \& Oh 2004; Di Matteo et al. 2005; Hopkins \& Elvis 2010) to efficiently drive AGN feedback.

From the spectral decomposition, we know that the Balmer decrement is $\rm H\alpha/H\beta  = 5.16\pm0.15$, however,
the intrinsic value of $\rm H\alpha/H\beta$ of the BLR is 3.06 with a standard deviation of 0.03 dex (Dong et al. 2008).
The observed Balmer decrement suggests that the BLR is reddened by $E(B-V)=0.57\pm0.03$ mag using the SMC-type extinction law,
which is smaller than the extinction obtained from the nuclear continuum.
This inference is consistent with the fact of the absorption gases partially obscuring the BLR.
That is why we use the $M_{\rm BH}-L_{\rm 5100}$ formalism rather than the $M_{\rm BH}-L_{\rm H\alpha}$ formalism in the estimation of the black hole mass.
Meanwhile, one can clearly find the absorption troughs of \ion{Ca}{2} H and K in the optical spectrum of \J1450.
In the right-third panel of Figure \ref{f3}, we also show the normalized absorption fluxes of \ion{Ca}{2} H, which are easily described by a single Gaussion profile with a blueshifted velocity of $309\pm33$ \mbox{km s$^{-1}$} and a width of $232\pm39$ \mbox{km s$^{-1}$}.
Obviously, \ion{Ca}{2} H and K absorption lines have different velocity structures with those of hydrogen Balmer and \ion{He}{1}*.
The above results rise up a question: ``which absorption system is the dust completely (or partly) associated with?''
Two factors of $E_{(B-V)}^{nuclear}$ and $E_{(B-V)}^{host}$ are set to represent the extinction associated with the absorption of hydrogen Balmer and \ion{He}{1}* (the gases in the nuclear region) or \ion{Ca}{2} H and K (the gases in the host galaxy),
which will meet the following conditions:\\
\begin{align}
E_{(B-V)}^{nuclear}+E_{(B-V)}^{host}=0.72\pm0.01 ~\rm mag,
\label{eq4}
\end{align}
\begin{align}
\dfrac{Cf~f_{\rm H\alpha}~A(E_{(B-V)}^{nuclear}+E_{(B-V)}^{host},\lambda)+(1-Cf)~f_{\rm H\alpha}~A(E_{(B-V)}^{host},\lambda)}{Cf~f_{\rm H\beta}~A(E_{(B-V)}^{nuclear}+E_{(B-V)}^{host},\lambda)+(1-Cf)~f_{\rm H\beta}~A(E_{(B-V)}^{host},\lambda)}=5.16\pm0.15.
\label{eq5}
\end{align}
Then the extinction in the nuclear region is $E_{(B-V)}^{nuclear}=0.21\pm0.04$ mag.
The dusty materials are mainly associated with \ion{Ca}{2} H and K absorption lines and exist in the host galaxy.
That implies that the dust-free model we adopted in the Cloudy simulation is approximately reasonable in a certain extent.

\section{Summary}
In this paper, we present detailed analysis of the SDSS-III/BOSS optical spectrum and newly obtained P200 TripleSpec NIR spectrum for \J1450.
The object is highly dust reddened with an extinction of $E(B-V) = 0.72\pm0.01$ mag under the SMC extinction law, which is dominant by the dusty materials associated with \ion{Ca}{2} H and K rather than the hydrogen Balmer and metastable helium absorption system.
\J1450\ is classified as the fifth Balmer BAL quasar based on the widths of Balmer (H$\alpha$, H$\beta$, H$\gamma$, and H$\delta$) absorption troughs.
Indeed, the troughs of H$\alpha$ and \ion{He}{1}* $\lambda10830$ are saturated, and the other unsaturated Balmer and \ion{He}{1}* $\lambda3889$ absorption lines are modeled by the two-Gaussian profiles with the center velocity shifts of $-931\pm33$ and $-499\pm39$ km s$^{-1}$ and the widths of $121\pm28$ and $196\pm37$ km s$^{-1}$, respectively.
The advantage of multiple lines of hydrogen Balmer and \ion{He}{1}* enables us to measure accurately the column density and covering factor of the lines.
And $\rm H(n = 2)$ is sensitive to the gas density, while \ion{He}{1}* is sensitive to the ionization state of the absorption gases.
The depths of hydrogen Balmer and \ion{He}{1}* $\lambda\lambda3889,10830$ absorption troughs suggest that the accretion disk is completely obscured by the outflow gases and the covering factor of the BLR is only $0.39\pm0.03$.
The total  column densities are $N_{\rm H(n=2)}=(4.89\pm1.07)\times10^{14}\rm~ cm^{-2}$ and $N_{\rm HeI*}=(1.01\pm0.19)\times10^{15}\rm~cm^{-2}$ by integrating the true optical depth.
Extensive photoionization models are calculated using Cloudy.
Together with the undetected \ion{Fe}{2}*, the physical parameters of the absorption gases are constrained to be
$10^{-1.4} \lesssim U \lesssim 10^{-0.8}$, $10^{8.2\pm0.4} \lesssim n_{\rm H} \lesssim 10^{9.0\pm0.4}$ cm$^{-3}$,
and $10^{22.0\pm0.2} \lesssim N_{\rm H} \lesssim 10^{22.2-22.3}$ cm$^{-2}$.
The absorption gases are estimated to be in the vicinity of the surface of the BLR  or at most 3.12 pc away from the central engine.

\acknowledgments{This work is supported by National Natural Science Foundation of China (NSFC-11573024, 11473025, 11421303) and National Basic Research Program of China (the 973 Program 2013CB834905). T. Ji is supported by National Natural Science Foundation of China  (NSFC-11503022) and Natural Science Foundation of Shanghai (NO. 15ZR1444200). P. Jiang is supported by  National Natural Science Foundation of China  (NSFC-11233002).
We acknowledge the use of the Hale 200-inch Telescope at Palomar Observatory through the Telescope Access Program (TAP), as well as the archive data from the SDSS, 2MASS and WISE Surveys. TAP is funded by the Strategic Priority Research Program. The Emergence of Cosmological Structures (XDB09000000), National Astronomical Observatories, Chinese Academy of Sciences, and the Special Fund for Astronomy from the Ministry of Finance. Observations obtained with the Hale Telescope at Palomar Observatory were obtained as part of an agreement between the National Astronomical Observatories, Chinese Academy of Sciences, and the California Institute of Technology. Funding for SDSS-III has been provided by the Alfred P. Sloan Foundation, the Participating Institutions, the National Science Foundation, and the U.S. Department of Energy Office of Science. The SDSS-III Web site is http:// www.sdss3.org/.}

\clearpage

\begin{deluxetable}{ccccc}
\tabletypesize{\scriptsize}
\tablewidth{0pt}
\tablenum{1}
\tablecaption{Photometric Observations
\label{tab1} }
\tablehead{
\colhead{Band}  & \colhead{Magnitude} & \colhead{Date of Observation}& \colhead{Facility} & \colhead{Reference}}
\startdata
$\textit{u}$          &$ 23.09\pm 0.47 $ &2002 Sep. 05&SDSS&1,2\\
$\textit{g}$          &$ 21.11\pm 0.04 $ &2002 Sep. 05&SDSS&1,2\\
$\textit{r}$          &$ 20.63\pm 0.04 $ &2002 Sep. 05&SDSS&1,2\\
$\textit{i}$          &$ 19.57\pm 0.03 $ &2002 Sep. 05&SDSS&1,2\\
$\textit{z}$          &$ 18.77\pm 0.05 $ &2002 Sep. 05&SDSS&1,2\\
$  J  $               &$ 16.66\pm 0.14 $ &1998 June 16&2MASS&3\\
$  H  $               &$ 16.12\pm 0.21 $ &1998 June 16&2MASS&3\\
$  K_s$               &$ 15.10\pm 0.18 $ &1998 June 16&2MASS&4\\
$  W1 $               &$ 13.53\pm 0.01 $ &2010 June 24&WISE&4\\
$  W2 $               &$ 12.18\pm 0.01 $ &2010 June 24&WISE&4\\
$  W3 $               &$  9.36\pm 0.03 $ &2010 June 24&WISE&4\\
$  W4 $               &$  7.15\pm 0.06 $ &2010 June 24&WISE&4
\enddata
\tablenotetext{References:}{(1) York et al. (2000); (2) Alam et al. (2015); (3) Skrutskie et al. (2006); (4) Wright et al. (2010)}
\end{deluxetable}

\figurenum{1}

\figurenum{2}

\figurenum{3}

\figurenum{4}

\end{document}